\documentstyle[12pt]{article}

\begin{document}
\baselineskip 0.7cm
\abovedisplayskip 5.7mm
\belowdisplayskip 5.7mm

\newcommand{\gsim}{ \mathop{}_{\textstyle \sim}^{\textstyle >} }
\newcommand{\lsim}{ \mathop{}_{\textstyle \sim}^{\textstyle <} }
\newcommand{\vev}[1]{ \left\langle {#1} \right\rangle }
\newcommand{\lsp}{ \left ( }
\newcommand{\rsp}{ \right ) }
\newcommand{\lmp}{ \left \{ }
\newcommand{\rmp}{ \right \} }
\newcommand{\llp}{ \left [ }
\newcommand{\rlp}{ \right ] }
\newcommand{\labs}{ \left | }
\newcommand{\rabs}{ \right | }
\newcommand{\KEV}{ {\rm keV} }
\newcommand{\MEV}{ {\rm MeV} }
\newcommand{\GEV}{ {\rm GeV} }
\newcommand{\TEV}{ {\rm TeV} }

\newcommand{\mgra}{ m_{3/2} }
\newcommand{\mh}{ M_{H} }

\renewcommand{\thefootnote}{\fnsymbol{footnote}}
\setcounter{footnote}{1}

\begin{titlepage}
\begin{flushright}
TU-455 \\
Mar. 1994
\end{flushright}

\vskip 0.35cm
\begin{center}
{\Large \bf A Solution to the Polonyi Problem in the Minimum SUSY-GUT}
\vskip 0.5cm

\vskip 0.25cm

T. Moroi
\footnote{Fellow of the Japan Society for the Promotion of Science.}
and T. Yanagida

\vskip 0.25cm
{\it Department of Physics, Tohoku University, Sendai 980, Japan}

\vskip 1.5cm

\abstract{
We show that the Polonyi problem is solved in the minimum SUSY-GUT
model in which a self-coupling strength for a heavy Higgs $\Sigma$,
$\lambda\Sigma^{3}$, is very small $\lambda\sim 10^{-6}$. It is stressed
that with this small $\lambda$ the mass of the physical $\Sigma$ becomes
$m_{\Sigma} \sim 10^{12}\GEV$ and the unification scale is raised up to
the gravitational one, $M\simeq 2\times 10^{18}\GEV$. A potential
problem, however, is also pointed out in this GUT model.
}

\vskip 10cm
(Submitted to {\it Prog. Theor. Phys.})

\end{center}
\end{titlepage}

\renewcommand{\thefootnote}{\arabic{footnote}}
\setcounter{footnote}{0}

The Polonyi problem\cite{PLB131-59} is one of serious problems in $N$=1
supergravity models in which the supersymmetry (SUSY) is spontaneously
broken in hidden sector\cite{PRep110-1}. In these theories there is
necessarily a light massive field (Polonyi field) $\phi$ of the mass
$m_{\phi}=O(\mgra)$\cite{PLB131-59,polonyi-plob}. The Polonyi field
$\phi$ couples only gravitationally to particles of observed sector and
hence it decays at a very late time.  Under quite general assumptions,
the coherent mode of $\phi$ dominates the energy density of the universe
until $T\sim 10^{-2}\MEV$ and its decay releases an unacceptable amount
of entropy\cite{PLB131-59}. No convincing mechanism has been found to
solve this serious problem.

In this letter we show that there is indeed no Polonyi problem in
supergravity models where a pair of light Higgs multiplets is obtained
by a fine tuning of parameters, just like in the minimum SUSY-SU(5)
model and the self-coupling $\lambda$ (defined in eq.(\ref{super2})) is
very small.

Let us start with the Polonyi superpotential\cite{polonyi}
\begin{eqnarray}
W(z) = \mu^{2} ( z + a )
\label{super}
\end{eqnarray}
with $a=(2-\sqrt{3})M$ and $M$ being the gravitational scale
$M=M_{planck}/\sqrt{8\pi}$ $\simeq 2.4 \times 10^{18} \GEV$. With the
K\"ahler potential $K(z, z^{*}) = z z^{*}$, the scalar potential is
given by
\begin{eqnarray}
V(z) =
\exp \lsp \frac{z z^{*}}{M^{2}} \rsp
\lmp \labs \frac{z^{*} W}{M^{2}} + \mu^{2} \rabs^{2} -
3 \frac{\labs W \rabs^{2}}{M^{2}} \rmp .
\label{pot}
\end{eqnarray}
This potential has a minimum at $\vev{z} = (\sqrt{3} - 1)M$ with a
vanishing cosmological constant $\Lambda_{cos}=0$. At the minimum
$z=\vev{z}$ the SUSY is broken, giving the gravitino mass
\begin{eqnarray}
\mgra =
\exp \lsp \frac{\vev{z} \vev{z^{*}}}{2 M^{2}} \rsp
\frac{\labs W(\vev{z}) \rabs}{M^{2}}
\simeq
\frac{\mu^{2}}{M} e^{2-\sqrt{3}}.
\label{mgra}
\end{eqnarray}
There is a flat direction called Polonyi field
$\phi_{0}$ that is defined as
\begin{eqnarray}
\phi_{0} \equiv z - \vev{z}.
\label{phi}
\end{eqnarray}
Its mass is at the same order of the gravitino mass,
\begin{eqnarray}
m_{\phi_{0}} \simeq \sqrt{2\sqrt{3}} \mgra.
\label{mphi}
\end{eqnarray}

We now introduce Higgs multiplets. In order to demonstrate our main
point clearly, we take all Higgs fields $\Sigma$, $H$ and $\bar{H}$ to
be singlets and assume the following superpotential;
\begin{eqnarray}
W =
\lambda
\lsp
\frac{1}{3} \Sigma_{0}^{3}
- \frac{1}{2} v \Sigma_{0}^{2} + \frac{1}{6} v^{3}
\rsp
+ g \bar{H} \Sigma_{0} H - \mh \bar{H} H +
W(z).
\label{super2}
\end{eqnarray}
In the global SUSY limit the $\Sigma_{0}$ has a vacuum-expectation value
$\vev{\Sigma_{0}}=v$. With a fine tuning $\mh\simeq gv$, we get a pair
of light Higgs multiplets whose mass is set as
\begin{eqnarray}
m_{H} = g \vev{\Sigma_{0}} - \mh \simeq O(\mgra).
\label{mh}
\end{eqnarray}
We choose $\mgra \simeq 1\TEV$ in this letter.

Total scalar potential is given by
\begin{eqnarray}
V &=&
\exp
\lsp
\frac{z z^{*} + \Sigma_{0} \Sigma_{0}^{*} + \cdot \cdot \cdot}{M^{2}}
\rsp
{\Biggl \{ }
\labs
\frac{\Sigma_{0}^{*} W}{M^{2}} + \frac{\partial W}{\partial \Sigma_{0}}
\rabs^{2} +
\labs
\frac{H^{*} W}{M^{2}} + \frac{\partial W}{\partial H}
\rabs^{2}
\nonumber \\
&&~~~~~~~~~~~~~~~~~ +
\labs
\frac{\bar{H}^{*} W}{M^{2}} + \frac{\partial W}{\partial \bar{H}}
\rabs^{2} +
\labs \frac{z^{*} W}{M^{2}} + \mu^{2} \rabs^{2} -
3 \frac{\labs W \rabs^{2}}{M^{2}}
{\Biggr \}} .
\label{full-pot}
\end{eqnarray}
An important observation is that the first term in the bracket possesses
a non-negligible mixing term between $\phi_{0}$ and $\Sigma_{0}$ fields.

To see this we neglect the exponential term in eq.(\ref{full-pot}) for
simplicity. The relevant part of $V$ is written as
\begin{eqnarray}
V &=&
\labs
\frac{\Sigma^{*}_{0}}{M^{2}} \mu^{2} \lsp \phi_{0} + M \rsp
+ \lambda \lsp \Sigma_{0}^{2} - v \Sigma_{0} \rsp
\rabs^{2}
\nonumber \\ && +
\labs
\frac{\phi_{0}^{*} + \vev{z^{*}}}{M^{2}} \mu^{2} \lsp \phi_{0}+M \rsp
+ \mu^{2}
\rabs^{2}
\nonumber \\ && -
\frac{3}{M^{2}}
\labs
\lambda \lsp \frac{1}{3} \Sigma_{0}^{3} - \frac{1}{2} v \Sigma_{0}^{2}
+ \frac{1}{6} v^{3} \rsp
+ \mu^{2} \lsp \phi_{0} + M \rsp
\rabs^{2}.
\label{pot2}
\end{eqnarray}
We see that the physical $\Sigma$ and the Polonyi field $\phi$ is
defined as
\begin{eqnarray}
\Sigma &\simeq& \Sigma_{0} + \frac{\mu^{2}}{\lambda M^{2}} \phi_{0},
\label{phys-s} \\
\phi &\simeq& \phi_{0} - \frac{\mu^{2}}{\lambda M^{2}} \Sigma_{0}.
\label{phys-p}
\end{eqnarray}

Substituting eq.(\ref{phys-s}) and eq.(\ref{phys-p}) into
eq.(\ref{pot2}), we obtain the potential $V$ as
\begin{eqnarray}
V &\simeq&
\lambda^{2} v^{2}
\labs \Sigma - \lsp v - \frac{\mu^{2}}{\lambda M} \rsp \rabs^{2}
\nonumber \\ &&
+ 2 \lsp \frac{\mu^{2}}{M} \rsp^{2} \labs \phi \rabs^{2}
+ \lsp \sqrt{3} - 1 \rsp \lsp \frac{\mu^{2}}{M} \rsp^{2}
\lsp \phi^{2} + \phi^{*~2} \rsp.
\label{pot3}
\end{eqnarray}
Here we have assumed all parameters are real. It is now clear that
vacuum-expectation values for the fields $\Sigma$ and $\phi$ are given
by
\begin{eqnarray}
\vev{\Sigma} &\simeq& v - \frac{\mu^{2}}{\lambda M},
\label{vev-s} \\
\vev{\phi} &\simeq& \frac{\mu^{2}}{2 \lambda M}.
\label{vev-p}
\end{eqnarray}
and $\phi$ is indeed the physical Polonyi field.  Then, the Polonyi
field $\phi$ couples to fermion partners of $H$ and $\bar{H}$ denoted by
$\psi_{H}$ and $\psi_{\bar{H}}$ through the above mixing (\ref{phys-s})
and (\ref{phys-p}). The effective Yukawa coupling of $\phi$ is given by
\begin{eqnarray}
{\cal L}_{yukawa} =
\lsp \frac{\mgra}{\lambda M} \rsp
g \phi \psi_{\bar{H}} \psi_{H} + h.c.
\label{yukawa}
\end{eqnarray}
Notice that if $\lambda$ is very small $\sim 10^{-6}$, this Yukawa
coupling is not negligible and induces a fast decay of $\phi$ enough to
solve the Polonyi problem as explained below.

The rate of the decay $\phi \rightarrow \psi_{H} + \psi_{\bar{H}}$ is
calculated as\footnote
{At the decay time the amplitude of $\phi$ is
\begin{eqnarray*}
\phi \sim \frac{\alpha}{4}
\lsp \frac{\mgra}{\lambda M} \rsp^{2} M.
\end{eqnarray*}
With this $\phi$ the Higgs fermion $\psi_{H}$ get a mass
\begin{eqnarray*}
m_{\psi_{H}} \simeq
\frac{g \alpha}{4}
\lsp \frac{\mgra}{\lambda M} \rsp^{3} M,~~
\lsp \alpha = \frac{g^{2}}{4 \pi} \rsp .
\end{eqnarray*}
Requiring the $2 m_{\psi_{H}} < m_{\phi}$ so that the decay of $\phi
\rightarrow \psi_{\bar{H}} + \psi_{H}$ is possible, we get a constraint
on $\lambda$ with $g\sim O(1)$
\begin{eqnarray*}
\lambda > \lsp \frac{\mgra}{M} \rsp^{2/3} \sim 10^{-10},
\end{eqnarray*}
for $\mgra = 1\TEV$.
}
\begin{eqnarray}
\Gamma_{\phi} \simeq \frac{\alpha}{4}
\lsp \frac{\mgra}{\lambda M} \rsp^{2}  m_{\phi},
\label{gamma-p}
\end{eqnarray}
and the reheating temperature is
\begin{eqnarray}
T_{R}
\simeq 0.1 \sqrt{\Gamma_{\phi} M_{planck}}
\simeq 0.1 \lsp \frac{\sqrt{\alpha}\mgra}{2\lambda M} \rsp
\sqrt{m_{\phi} M_{planck}}.
\label{tr}
\end{eqnarray}
If one requires $T_{R} \gsim O(100\GEV)$ so that the electroweak
baryogenesis is possible\cite{PLB155-36}, one gets for $\alpha \sim
O(1)$ and $\mgra \simeq 1\TEV$
\begin{eqnarray}
\lambda \lsim 10^{-6}.
\label{lambda}
\end{eqnarray}
However, we do not need to create baryon asymmetry at the electroweak
scale, since the dilution factor $D$ of the baryon asymmetry from the
$\phi$ decay is only\footnote
{In the previous analysis\cite{PLB131-59}, the decay rate of $\phi$ is
assumed as $\Gamma_{\phi}\simeq m_{\phi}^{3}/M_{planck}^{2}$ which leads
to the dilution factor $D\simeq 10^{-14}$.}
\begin{eqnarray}
D \simeq 10^{-6},
\end{eqnarray}
with $\lambda \simeq 10^{-6}$ and the initial value $\phi\simeq M$. This
requires the primordial baryon asymmetry to be $\Delta B/s\sim 10^{-4} -
10^{-5}$ which may be produced in a much early epoch of the universe.

It is a straightforward task to incorporate our general mechanism into
the minimum SUSY-SU(5) model. The superpotential is given by
\begin{eqnarray}
W &=&
\frac{1}{3} \lambda {\rm tr} \Sigma^{3} +
\frac{1}{2} \lambda v {\rm tr} \Sigma^{2} -
5 \lambda v^{3}
\nonumber \\ &&
+ g \bar{H}^{\alpha} \Sigma^{\beta}_{\alpha} H_{\beta} +
\mh \bar{H}^{\alpha} H_{\alpha} +
W(z),
\label{su5}
\end{eqnarray}
where $\Sigma$, $H$ and $\bar{H}$ are {\bf 24}, {\bf 5} and ${\bf
5^{*}}$ of SU(5). In the global SUSY limit the $\Sigma$ field has a
vacuum-expectation value
\begin{eqnarray}
\vev{\Sigma} =  v
\lsp
\begin{array}{ccccc}
2 & & & & \\
  & 2 & & & \\
  & & 2 & & \\
  & & & -3 & \\
  & & & & -3
\end{array}
\rsp ,
\label{vev}
\end{eqnarray}
and masses of triplet and doublet Higgs multiplets $H_{c}$ and $H_{f}$
are
\begin{eqnarray}
m_{H_{c}} &=& 2 g v + \mh,
\\
m_{H_{f}} &=& -3 g v + \mh.
\label{flavor}
\end{eqnarray}
By a fine tuning $\mh \simeq 3gv$, we obtain a pair of light Higgs
doublets $H_{f}$ and $\bar{H}_{f}$ and a pair of heavy Higgs triplets
$H_{c}$ and $\bar{H}_{c}$. We have checked that the similar mixings
between $\Sigma$ and $\phi$ takes place in the SU(5) broken phase as in
the singlet model, inducing a Yukawa coupling of $\phi$;
\begin{eqnarray}
{\cal L}_{yukawa} =
3 \lsp \frac{\mgra}{\lambda M} \rsp
g \phi \psi_{\bar{H}_{f}} \psi_{H_{f}} + h.c.
\label{yukawa-GUT}
\end{eqnarray}

Finally, we stress that the small self-coupling $\lambda \simeq 10^{-6}$
suggests that the mass of physical $\Sigma$ ({\bf 24}) in SUSY SU(5)
model is relatively small $m_{\Sigma}\sim\lambda\vev{\Sigma}(\sim
O(10^{12}\GEV))$. As pointed out by Hisano, Murayama and
Yanagida\cite{NPB402-46}, this is still consistent with the unification
of three gauge coupling constants if the GUT scale $\vev{\Sigma} \sim
O(10^{18}\GEV)$. (This may be rather interesting for superstring
theories.) The crucial point in this letter is that the Polonyi field
$\phi$ couples to $\psi_{H}$ and $\psi_{\bar{H}}$ with the strength
$(\mgra/m_{\Sigma})$ as shown in eq.(\ref{yukawa}) which arises from the
mixing between the $\phi$ and $\Sigma$ fields. The coupling of $\phi$ is
no longer suppressed by $(1/M)$ as expected previously.  However, there
is a potential problem in this GUT model. As seen in eq.(\ref{vev-s}),
the shift of $\vev{\Sigma}$ is not small enough to maintain the
hierarchy achieved in the global SUSY model. Therefore, we must rechoose
$M_{H}$ so that the physical $m_{H_{f}}$ is $O(\mgra)$ after the shift
of $\vev{\Sigma}$.  Moreover, the shift also produces a large soft-SUSY
breaking term $\sqrt{3}g\frac{\mu^{4}}{\lambda M^{2}}
\bar{H}H$. The coefficient of this dangerous term, however, depends
strongly on the form of K\"ahler potential and in fact we have found a
non-minimum K\"ahler potential with which such a soft-SUSY breaking term
vanishes.\footnote
{In this model we take the K\"ahler potential $K=z^{*}z\exp
(\Sigma^{*}\Sigma/M^{2})+\Sigma^{*}\Sigma$ and the superpotential for
$z$, $W(z)=\mu^{2}M+c_{1}(z/M)+c_{2}(z/M)^{2}+\cdot\cdot\cdot$. The
coefficients $c_{i}$ are fixed from the requirement that the scalar
potential $V(z,\Sigma)$ has a minimum at $\vev{z}=0$ with a vanishing
cosmological constant.}
Since more fine tuning is required, the GUT
model with small $\lambda$ is less attractive.  Nevertheless, it is very
much encouraging that a fine tuning for guaranteeing the large mass
hierarchy in Higgs sector solves automatically the serious cosmological
problem.

We thank J.~Hisano and M.~Yamaguchi for a useful discussion.

\newpage

\newcommand{\Journal}[4]{{\sl #1} {\bf #2} {(#3)} {#4}}
\newcommand{\PL}{Phys. Lett.}
\newcommand{\PR}{Phys. Rev.}
\newcommand{\PRL}{Phys. Rev. Lett.}
\newcommand{\NP}{Nucl. Phys.}
\newcommand{\PTP}{Prog. Theor. Phys.}


\begin{thebibliography}{99}

\bibitem{PLB131-59}
G.D.~Coughlan, N.~Fischler, E.W.~Kolb, S.~Raby and G.G.~Ross,
{\sl \PL} {\bf B131} (1983) 59.

\bibitem{PRep110-1}
For a review, H.P.~Nilles,
{\sl Phys. Rep.} {\bf 110} (1984) 1.

\bibitem{polonyi-plob}
T.~Banks, D.B. Kaplan and A.E. Nelson,
preprint UCSD-PTH-93-26 (1993); \\
B.~de~Carlos, J.A.~Carlos, F. Quevedo and E.~Roulet,
{\sl \PL} {\bf B318} (1993) 447.

\bibitem{polonyi}
J.~Polonyi,
Budapest preprint KFK-1977-93 (unpublished).

\bibitem{PLB155-36}
V.A.~Kuzmin, V.A.~Rubakov and M.E.~Shaposhnikov,
{\sl \PL} {\bf B155} (1985) 36.

\bibitem{NPB402-46}
J.~Hisano, H.~Murayama and T.~Yanagida,
{\sl \NP} {\bf B402} (1993) 46.

\end{thebibliography}
\end{document}